\documentclass[12pt,a4paper]{article}
\usepackage{amsmath}
\usepackage{amssymb}
\usepackage{bm}% bold math
\usepackage{epsfig, graphicx, euscript}
\newcommand{\be}{\begin{equation}}
\newcommand{\ee}{\end{equation}}
\newcommand{\bea}{\begin{eqnarray}}
\newcommand{\eea}{\end{eqnarray}}

\newcommand{\mbss}[1]{_{\mbox{\scriptsize #1}}}

\newcommand{\mbsu}[1]{\mbox{\scriptsize #1}}

\begin{document}
\begin{center}
\bf
CALCULATIONS OF THE ISOSCALAR GIANT MONOPOLE RESONANCE
WITHIN THE SELF-CONSISTENT RPA\\
AND ITS EXTENSIONS
\end{center}

\begin{center}
V.~Tselyaev$^1$,
S.~Krewald$^2$,
E.~Litvinova$^{3,4,5}$,
J.~Speth$^2$\\
\end{center}

\begin{center}
\it
$^1$Nuclear Physics Department, V. A. Fock Institute of Physics,\\
St. Petersburg State University, 198504 St. Petersburg, Russia\\
$^2$Institut f\"ur Kernphysik, Forschungszentrum J\"ulich,
52425 J\"ulich, Germany\\
$^3$Institute of Physics and Power Engineering,
249033 Obninsk, Russia\\
$^4$Gesellschaft f\"ur Schwerionenforschung mbH,
64291 Darmstadt, Germany\\
$^5$Institut f\"{u}r Theoretische Physik,
Goethe-Universit\"{a}t, 60438 Frankfurt am Main, Germany
\end{center}
\bigskip

{
%\small
\footnotesize
The strength distributions of the isoscalar giant monopole resonance
have been calculated in
$^{16}$O, $^{40}$Ca, $^{90}$Zr, $^{112-124}$Sn, $^{144}$Sm,
and $^{208}$Pb nuclei
within the self-consistent random phase approximation and its extensions
which include pairing correlations and quasiparticle-phonon coupling.
The results are compared with the available experimental data.
The problem of the nuclear matter incompressibility is discussed.
}
\bigskip
\bigskip

Theoretical description of the isoscalar giant monopole resonance (ISGMR)
is an important issue, first of all, because its energy
is closely related to the value of the incompressibility modulus
of infinite nuclear matter (INM) $K_{\infty}$ (see \cite{BLM79,B80}),
which in turn is a universal characteristic of the
effective nuclear forces.
The most widely used method of evaluating $K_{\infty}$ is based on
the self-consistent Hartree-Fock (HF) or random phase approximation (RPA)
calculations of the mean energies of the ISGMR
using effective Skyrme or Gogny forces
(see, e.g., Refs.
\cite{BLM79,B80,BBDG95,CGMBB,SYZGZ}).
Because $K_{\infty}$ can be calculated from the known parameters
of the given force, its value is estimated as the one corresponding
to the force that gives the best description of the experimental
energies. The non-relativistic estimates obtained
in such a way lead to the value $K_{\infty} = 210 \pm 30$ MeV
(see, e.g., Refs. \cite{B80,BBDG95}),
though the recent results testify to the upper limit of this estimate
(see \cite{CGMBB,SYZGZ}).

However, neither the HF approximation nor the RPA can
provide full description of the experimental data. In particular,
they cannot reproduce the total width $\Gamma$ of the giant resonance,
since in these models the mechanism responsible for the formation
of the spreading width $\Gamma^{\downarrow}$ is absent.
At the same time, it is well known that the spreading width
is a considerable part of $\Gamma$.
In the present work we report the results of the ISGMR calculations
within three models: (i) the RPA; (ii) the quasiparticle RPA (QRPA);
(iii) the quasiparticle time blocking approximation
(QTBA, see \cite{QTBA1,QTBA2}) which is
an extension of the QRPA including quasiparticle-phonon coupling (QPC).
The QPC provides the necessary mechanism producing the spreading width
in the QTBA.
The pairing correlations are taken into account both in the QRPA and
in the QTBA.

The basic equation of our approach is the equation for the effective
response function $R^{\mbsu{eff}}(\omega)$.
It has the same form both in the (Q)RPA and in the QTBA.
In the shorthand notations it reads
(we will follow notations of Ref.~\cite{QTBA2})
\be
R^{\mbsu{eff}}(\omega) = A(\omega) - A(\omega)\,{\cal F}\,
R^{\mbsu{eff}}(\omega)\,,
\label{bse1}
\ee
where $A(\omega)$ is a correlated propagator
and $\cal F$ is an amplitude of the effective residual interaction.
In the present work we use the version of the QTBA in which
the ground-state correlations caused by the QPC are neglected.
In this case the correlated propagator $A(\omega)$ is defined
by the equation
\be
A(\omega) = \tilde{A}(\omega)
- \tilde{A}(\omega)\,\bar{\Phi}(\omega)\,A(\omega)\,,
\label{cprp}
\ee
where $\tilde{A}(\omega)$ is the uncorrelated QRPA propagator,
\be
\bar{\Phi}(\omega) =
\Phi^{(\mbsu{res})}(\omega) - \Phi^{(\mbsu{res})}(0)\,,
\label{bphi}
\ee
and $\Phi^{(\mbsu{res})}(\omega)$ is a resonant part of the
interaction amplitude responsible for the QPC in our model
(see Refs.~\cite{QTBA1,QTBA2} for details).
In the case of the (Q)RPA, the amplitude $\Phi^{(\mbsu{res})}(\omega)$
is set equal to zero, and $A(\omega)$ reduces to the uncorrelated
propagator $\tilde{A}(\omega)$.

The dynamical pairing effects (particle-particle channel contributions)
are taken into account both in the QRPA and in the QTBA.
This enables one to eliminate the $0^+$ spurious states on the QRPA level.
The problem of the $0^+$ spurious states in the QTBA is solved with
the help of a combination of the so-called subtraction procedure
(see Eq.~(\ref{bphi}) and Refs. \cite{QTBA1,QTBA2} for details)
and the projection technique described in \cite{ghst}.

The strength function of the ISGMR $S(E)$ is determined by
$R^{\mbsu{eff}}(\omega)$ via the formulas
\be
S(E) = -\frac{1}{\pi}\,\mbox{Im}\,\Pi (E + i \Delta)\,,
\label{defsf}
\ee
\be
\Pi (\omega) = -\frac{1}{2}\,\mbox{Tr}\,\bigl(
(eV^{\,0})^{\dag}\,R^{\,\mbsu{eff}}(\omega)\,(eV^{\,0})\bigr)\,,
\label{defpol}
\ee
where $\Pi (\omega)$ is the nuclear polarizability,
$E$ is an excitation energy,
$\Delta$ is a smearing parameter, $V^0$ is an external field,
and $e$ is an effective charge operator.
In the case of the isoscalar $0^+$ excitations the one-body operator
$eV^0$ is proportional to the identity matrices
both in the spin and in the isospin indices.
Its radial dependence is taken in our calculations in the form
$eV^0=r^2$.
The smearing parameter is taken to be equal to 500~keV
in all the calculations presented here.

We used a self-consistent calculation scheme based on the
HF and Bardeen-Cooper-Schrieffer (HF+BCS) approximation.
The details of this scheme are described in Ref.~\cite{gmrQTBA}.
Within the RPA the self-consistency is full.
The single-particle continuum is also exactly included
on the RPA level.
The self-consistent mean field and the effective interaction
(including the spin-orbital and the Coulomb contributions
in both quantities) are derived from the Skyrme energy functional.
In the calculations, two Skyrme force parametrizations
with the nucleon effective mass $m^* = m$ were used
(see Ref.~\cite{TBFP}):
T5 force with $K_{\infty}=202$ MeV and
T6 force with $K_{\infty}=236$ MeV.
The ISGMR calculations were performed in $^{16}$O, $^{40}$Ca, and $^{90}$Zr
nuclei, in the chain of the Sn isotopes, and in $^{144}$Sm and $^{208}$Pb
nuclei.

\begin{figure}
\begin{center}
\includegraphics*[scale=0.7,angle=0]{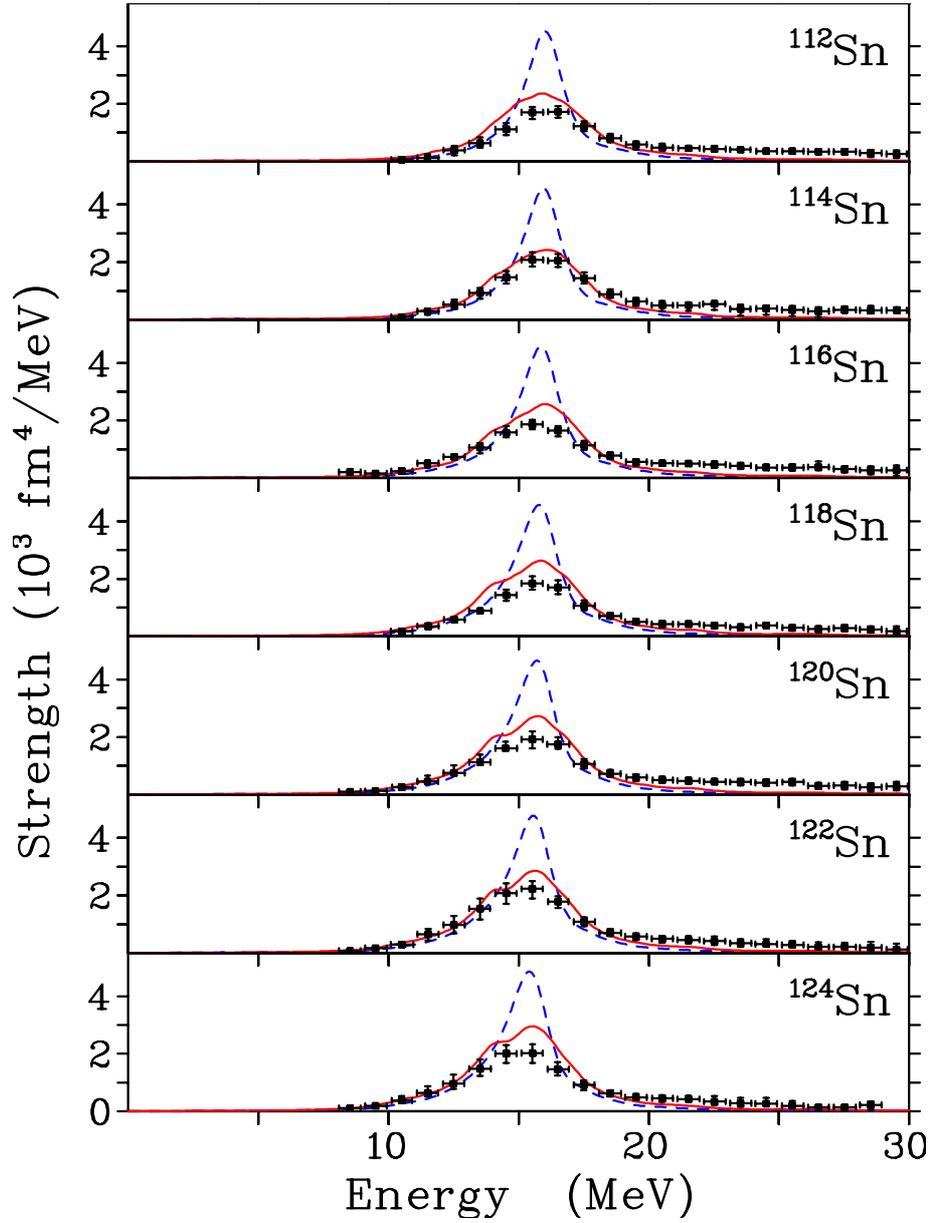}
\caption{\label{sn7is}
Isoscalar giant monopole resonance in the even-$A$ $^{112-124}$Sn isotopes
calculated within the QRPA (dashed line) and the QTBA (solid line)
using T5 Skyrme force.
The smearing parameter $\Delta$ is equal to 500 keV.
Experimental data (solid squares) are taken from Ref.~\cite{GMRexp}.}
\end{center}
\end{figure}

The results for the ISGMR strength distributions in the even-$A$
$^{112-124}$Sn isotopes are presented in Fig.~\ref{sn7is}
and in Table~\ref{tab1}.
The mean energies of the ISGMR drawn in the tables are defined via
the ratios of the energy-weighted moments $m_k$ determined as
\be
m_k = \int^{E_2}_{E_1} E^k S(E)\,d E\,.
\label{defmk}
\ee
\begin{table}
\begin{center}
\caption{\label{tab1}
Mean energies and Lorentzian-fit parameters for the ISGMR
strength distributions in the even-$A$ $^{112-124}$Sn isotopes.
Theoretical results are obtained within the QRPA and the QTBA
using T5 and T6 Skyrme forces (indicated in parentheses).
The mean energies are calculated for 10.5--20.5 MeV energy interval.
Experimental values are taken from Ref.~\cite{GMRexp}
(RCNP, Osaka University).}
\begin{tabular}{clccccc}
\hline
\hline
$\vphantom{\frac{A^{B^C}}{B}}$
& Method & $\sqrt{m_1/m_{-1}}$ & $m_1/m_0$
& $\sqrt{m_3/m_1}$ & $E_{\mbss{GMR}}$ & $\Gamma$ \\
&& (MeV) & (MeV) & (MeV) & (MeV) & (MeV) \\
\hline
\hline
$\vphantom{\frac{A^{B^C}}{B}}$
$^{112}$Sn & QRPA (T6) & 17.0 & 17.1 & 17.3 & 17.3 & 1.9 \\
           & QRPA (T5) & 15.8 & 15.9 & 16.1 & 15.9 & 1.8 \\
           & QTBA (T5) & 15.7 & 15.8 & 16.2 & 15.8 & 3.7 \\
           & Exp.
           & \; 16.1$^{+0.1}_{-0.1}$ \;
           & \; 16.2$^{+0.1}_{-0.1}$ \;
           & \; 16.7$^{+0.2}_{-0.2}$ \;
           & \; 16.1$^{+0.1}_{-0.1}$ \;
           & \, 4.0$^{+0.4}_{-0.4}$ \\
\hline
$\vphantom{\frac{A^{B^C}}{B}}$
$^{114}$Sn & QRPA (T6) & 16.9 & 17.0 & 17.2 & 17.3 & 2.0 \\
           & QRPA (T5) & 15.7 & 15.8 & 16.0 & 15.8 & 1.8 \\
           & QTBA (T5) & 15.6 & 15.7 & 16.1 & 15.7 & 3.7 \\
           & Exp.
           & 15.9$^{+0.1}_{-0.1}$
           & 16.1$^{+0.1}_{-0.1}$
           & 16.5$^{+0.2}_{-0.2}$
           & 15.9$^{+0.1}_{-0.1}$
           &  4.1$^{+0.4}_{-0.4}$ \\
\hline
$\vphantom{\frac{A^{B^C}}{B}}$
$^{116}$Sn & QRPA (T6) & 16.8 & 16.9 & 17.1 & 17.2 & 2.1 \\
           & QRPA (T5) & 15.6 & 15.6 & 15.9 & 15.7 & 1.9 \\
           & QTBA (T5) & 15.5 & 15.6 & 16.0 & 15.6 & 3.8 \\
           & Exp.
           & 15.7$^{+0.1}_{-0.1}$
           & 15.8$^{+0.1}_{-0.1}$
           & 16.3$^{+0.2}_{-0.2}$
           & 15.8$^{+0.1}_{-0.1}$
           &  4.1$^{+0.3}_{-0.3}$ \\
\hline
$\vphantom{\frac{A^{B^C}}{B}}$
$^{118}$Sn & QRPA (T6) & 16.6 & 16.7 & 17.0 & 17.1 & 2.1 \\
           & QRPA (T5) & 15.4 & 15.5 & 15.8 & 15.6 & 2.0 \\
           & QTBA (T5) & 15.4 & 15.5 & 15.9 & 15.5 & 3.9 \\
           & Exp.
           & 15.6$^{+0.1}_{-0.1}$
           & 15.8$^{+0.1}_{-0.1}$
           & 16.3$^{+0.1}_{-0.1}$
           & 15.6$^{+0.1}_{-0.1}$
           &  4.3$^{+0.4}_{-0.4}$ \\
\hline
$\vphantom{\frac{A^{B^C}}{B}}$
$^{120}$Sn & QRPA (T6) & 16.5 & 16.6 & 16.9 & 17.0 & 2.2 \\
           & QRPA (T5) & 15.3 & 15.4 & 15.7 & 15.5 & 2.1 \\
           & QTBA (T5) & 15.3 & 15.4 & 15.8 & 15.3 & 3.9 \\
           & Exp.
           & 15.5$^{+0.1}_{-0.1}$
           & 15.7$^{+0.1}_{-0.1}$
           & 16.2$^{+0.2}_{-0.2}$
           & 15.4$^{+0.2}_{-0.2}$
           &  4.9$^{+0.5}_{-0.5}$ \\
\hline
$\vphantom{\frac{A^{B^C}}{B}}$
$^{122}$Sn & QRPA (T6) & 16.4 & 16.5 & 16.8 & 16.9 & 2.3 \\
           & QRPA (T5) & 15.2 & 15.3 & 15.5 & 15.4 & 2.1 \\
           & QTBA (T5) & 15.1 & 15.3 & 15.7 & 15.2 & 3.8 \\
           & Exp.
           & 15.2$^{+0.1}_{-0.1}$
           & 15.4$^{+0.1}_{-0.1}$
           & 15.9$^{+0.2}_{-0.2}$
           & 15.0$^{+0.2}_{-0.2}$
           &  4.4$^{+0.4}_{-0.4}$ \\
\hline
$\vphantom{\frac{A^{B^C}}{B}}$
$^{124}$Sn & QRPA (T6) & 16.2 & 16.4 & 16.7 & 16.7 & 2.3 \\
           & QRPA (T5) & 15.0 & 15.1 & 15.4 & 15.2 & 2.2 \\
           & QTBA (T5) & 15.0 & 15.2 & 15.5 & 15.1 & 3.8 \\
           & Exp.
           & 15.1$^{+0.1}_{-0.1}$
           & 15.3$^{+0.1}_{-0.1}$
           & 15.8$^{+0.1}_{-0.1}$
           & 14.8$^{+0.2}_{-0.2}$
           &  4.5$^{+0.5}_{-0.5}$
$\vphantom{\frac{A}{A^{B^C}}}$ \\
\hline
\hline
\end{tabular}
\end{center}
\end{table}
\noindent
The energy interval limited by $E_1=10.5$ MeV and $E_2=20.5$ MeV
was taken the same as in Ref.~\cite{GMRexp},
where the experimental data on the strength distributions of the ISGMR
in the tin isotopes were reported.
The peak energies $E_{\mbss{GMR}}$ and the widths $\Gamma$ of the ISGMR
were obtained from the Lorentzian fit of the calculated functions $S(E)$.
As can be seen from the Table~\ref{tab1},
the agreement of the theoretical results with the experimental
mean and peak energies in the case of the T5 Skyrme force
is fairly good both in the QRPA and in the QTBA.
The fact that the mean and peak energies obtained
in the QRPA and in the QTBA are very close to each other
is explained by the subtraction procedure used in our calculations
(see Eq.~(\ref{bphi}) and Ref.~\cite{QTBA2} for the discussion).
The main reason of the agreement with the experiment in this case is
comparatively low value of the incompressibility modulus of INM
($K_{\infty}=202$ MeV) produced by the T5 Skyrme-force parametrization.
The other parametrizations with
$K_{\infty}$ around 240 MeV give too large mean energies of the ISGMR
in the considered tin isotopes as compared with the experiment.

For comparison, in Table~\ref{tab1} we draw the QRPA results
obtained with the T6 Skyrme force ($K_{\infty}=236$ MeV).
As can be seen, the T6 peak energies $E_{\mbss{GMR}}$ are greater than
the experimental values for the tin isotopes by 1.2--1.9 MeV.
This fact agrees with the results of Ref.~\cite{P07} where
it was obtained that the relativistic RPA calculations
based on the force with $K_{\infty}=230$ MeV consistently overestimate
the centroid energies of the ISGMR in the same tin isotopes.
\begin{table}[!ht]
\begin{center}
\caption{\label{tab2}
The same as in Table \ref{tab1}, but for
$^{16}$O, $^{40}$Ca, $^{90}$Zr, $^{144}$Sm, and $^{208}$Pb nuclei.
The theoretical mean energies are obtained for the following
energy intervals:
11--40 MeV in $^{16}$O, 10--55 MeV in $^{40}$Ca, and
5--25 MeV in $^{90}$Zr, $^{144}$Sm, and $^{208}$Pb.
Experimental data are taken from
Refs. \cite{isgmr16o,isgmr40ca,YCL99,YLC04}
(TAMU, Texas A\&M University)
and \cite{USI04,ISU03} (RCNP, Osaka University).
The data from Ref.~\cite{YCL99} for $^{90}$Zr denoted as
\cite{YCL99}$^{\,a}$ and \cite{YCL99}$^{\,b}$
correspond to the slice analysis and to the Gaussian fit, respectively.
}
\vspace{0.5em}
\begin{tabular}{clccccc}
\hline
\hline
$\vphantom{\frac{A^{B^C}}{B}}$
& Method & $\sqrt{m_1/m_{-1}}$ & $m_1/m_0$
& $\sqrt{m_3/m_1}$ & $E_{\mbss{GMR}}$ & $\Gamma$ \\
&& (MeV) & (MeV) & (MeV) & (MeV) & (MeV) \\
\hline
\hline
$\vphantom{\frac{A^{B^C}}{B}}$
$^{16}$O  & RPA $\;\;$ (T6) & 22.7 & 23.3 & 25.1 & 22.0 & 10.3 \\
          & RPA $\;\;$ (T5) & 21.3 & 21.8 & 23.4 & 20.5 &  7.2 \\
$\vphantom{\frac{A}{A^{B^C}}}$
          & Exp. $\;\;$ \cite{isgmr16o}
& 19.63$^{+0.38}_{-0.38}$
& 21.13$^{+0.49}_{-0.49}$
& 24.89$^{+0.59}_{-0.59}$ &
&  8.76$^{+1.82}_{-1.82}$ \\
\hline
$\vphantom{\frac{A^{B^C}}{B}}$
$^{40}$Ca & RPA $\;\;$ (T6) & 21.2 & 21.6 & 23.0 & 21.6 & 4.8 \\
          & RPA $\;\;$ (T5) & 19.7 & 19.9 & 21.1 & 19.7 & 4.0 \\
$\vphantom{\frac{A}{A^{B^C}}}$
          & Exp. $\;\;$ \cite{isgmr40ca}
&
& 19.18$^{+0.37}_{-0.37}$
&&
&  4.88$^{+0.57}_{-0.57}$ \\
\hline
$\vphantom{\frac{A^{B^C}}{B}}$
 $^{90}$Zr & QRPA (T6) & 18.0 & 18.2 & 18.6 & 18.0 & 3.0 \\
           & QTBA (T6) & 18.0 & 18.2 & 18.7 & 18.0 & 3.8 \\
           & QRPA (T5) & 16.6 & 16.8 & 17.2 & 16.5 & 2.0 \\
           & QTBA (T5) & 16.6 & 16.8 & 17.3 & 16.5 & 2.6 \\
           & Exp. $\;\;$ \cite{YCL99}$^{\,a}$
           & 17.81$^{+0.35}_{-0.35}$
           & 17.89$^{+0.20}_{-0.20}$ &&& \\
           & Exp. $\;\;$ \cite{YCL99}$^{\,b}$ &
           & 16.80 &&& \\
$\vphantom{\frac{A}{A^{B^C}}}$
           & Exp. $\;\;$ \cite{USI04}
           &&&& 16.6$^{+0.1}_{-0.1}$ & 4.9$^{+0.2}_{-0.2}$ \\
\hline
$\vphantom{\frac{A^{B^C}}{B}}$
$^{144}$Sm & QRPA (T6) & 15.8 & 16.0 & 16.4 & 15.8 & 2.0 \\
           & QTBA (T6) & 15.8 & 16.1 & 16.7 & 15.8 & 3.0 \\
           & QRPA (T5) & 14.6 & 14.7 & 15.1 & 14.5 & 1.5 \\
           & QTBA (T5) & 14.6 & 14.8 & 15.4 & 14.4 & 2.5 \\
           & Exp. $\;\;$ \cite{YLC04}
           && 15.40$^{+0.30}_{-0.30}$ &&& 3.40$^{+0.20}_{-0.20}$ \\
$\vphantom{\frac{A}{A^{B^C}}}$
           & Exp. $\;\;$ \cite{ISU03}
           &&&& 15.30$^{+0.11}_{-0.12}$ & 3.71$^{+0.12}_{-0.63}$ \\
\hline
$\vphantom{\frac{A^{B^C}}{B}}$
$^{208}$Pb & RPA $\;\;$ (T6) & 13.8 & 14.0 & 14.5 & 13.9 & 1.9 \\
           & QTBA (T6) & 13.8 & 14.1 & 14.8 & 13.9 & 3.2 \\
           & RPA $\;\;$ (T5) & 12.6 & 12.7 & 13.2 & 12.6 & 1.6 \\
           & QTBA (T5) & 12.6 & 12.8 & 13.5 & 12.5 & 2.8 \\
           & Exp. $\;\;$ \cite{YLC04}
           && 13.96$^{+0.20}_{-0.20}$ &&& 2.88$^{+0.20}_{-0.20}$ \\
$\vphantom{\frac{A}{A^{B^C}}}$
           & Exp. $\;\;$ \cite{USI04}
           &&&& 13.4$^{+0.2}_{-0.2}$ & 4.0$^{+0.4}_{-0.4}$ \\
\hline
\hline
\end{tabular}
\end{center}
\end{table}

In Table~\ref{tab2}, the (Q)RPA and QTBA results for
$^{16}$O, $^{40}$Ca, $^{90}$Zr, $^{144}$Sm, and $^{208}$Pb nuclei
are listed in comparison with the available ISGMR data.
As can be seen from this table,
the T5 Skyrme force leads to a better description of the experiment
in the light nuclei $^{16}$O and $^{40}$Ca.
The experimental value of the ISGMR peak energy in $^{90}$Zr
obtained by the RCNP (Osaka University, Ref. \cite{USI04})
and the value $m_1/m_0$ for this nucleus obtained by the Gaussian fit
in the TAMU experiment (Texas A\&M University, Ref. \cite{YCL99})
are also
fairly well reproduced by the calculations with the T5 Skyrme force.
The other experimental data for $^{90}$Zr and $^{144}$Sm nuclei
lie between the values calculated with the T5 and T6 forces.

It is worth noting that the value $K_{\infty}=202$ MeV corresponding to
the T5 Skyrme force lies within the interval
$210 \pm 30$ MeV which was considered for a long time
as the non-relativistic estimate for this quantity.
The recent results \cite{CGMBB,SYZGZ} inferring $K_{\infty}$ to be
230--240 MeV were obtained within the RPA and the constrained HF method
on the base of the experimental data in fact only for the one nucleus
$^{208}$Pb.
The mean energy $m_1/m_0$ of the ISGMR in $^{208}$Pb obtained
in the TAMU experiment \cite{YLC04}
is also nicely reproduced in our calculations
when the T6 force with $K_{\infty}=236$ MeV is used.
On the other hand, the T5 force gives the QTBA value $m_1/m_0$
for $^{208}$Pb which is lesser by 1.2~MeV as compared with this experiment.
Note, however, that the ISGMR data even for the well-studied nucleus
$^{208}$Pb are not so far quite unambiguous.
In particular, the experimental value of the ISGMR peak energy
in $^{208}$Pb measured in the RCNP experiment \cite{USI04}
is lesser by 0.6~MeV as compared with the value $m_1/m_0$
from Ref.~\cite{YLC04} and
lies between the QTBA values of $E_{\mbss{GMR}}$ obtained with
the T5 and T6 Skyrme forces.

In contrast to the mean energies, the (Q)RPA and the QTBA give substantially
different values for the total width $\Gamma$ of the ISGMR.
As was mentioned above, the reason is that the (Q)RPA does not produce
the spreading width $\Gamma^{\downarrow}$,
whereas in the QTBA it is formed due to the QPC effects.
The fact that $\Gamma^{\downarrow}$ is a considerable part of $\Gamma$
is illustrated by the results listed in the tables:
the (Q)RPA strongly underestimates the experimental values of the total width,
while reasonably good agreement is achieved in the QTBA.
The exception is the case of the light nuclei in which the resonance width
is fairly well reproduced already within the RPA.

In conclusion, we have presented the results of the theoretical analysis
of the ISGMR strength distributions in
$^{16}$O, $^{40}$Ca, $^{90}$Zr, $^{112-124}$Sn, $^{144}$Sm,
and $^{208}$Pb nuclei.
The calculations were performed within the (Q)RPA and the QTBA
which is an extension of the QRPA including
quasiparticle-phonon coupling (QPC).
We used calculational scheme based on the HF+BCS approximation
which is fully self-consistent on the RPA level.
In the calculations, two Skyrme force parametrizations were used.
The T5 parametrization
with comparatively low value of the incompressibility modulus
of infinite nuclear matter ($K_{\infty}=202$ MeV) allowed us
to achieve good agreement with the experimental data
for tin isotopes within the QTBA including resonance widths.
However, this parametrization underestimates
the experimental ISGMR mean energy for the $^{208}$Pb nucleus
which is usually used in the fit of the Skyrme force parameters.
On the other hand, the T6 Skyrme force with $K_{\infty}=236$ MeV
nicely reproduces the ISGMR mean energy for $^{208}$Pb but overestimates
the energies for $^{112-124}$Sn isotopes by more than one MeV.
For the light nuclei $^{16}$O and $^{40}$Ca, the T5 Skyrme force
gives a better description of the experimental ISGMR mean energies
as compared with the T6 force.
The experimental data on the ISGMR energies in $^{90}$Zr and $^{144}$Sm
nuclei lie between the values calculated by us with the T5 and T6 forces.
On the whole, these results do not allow us to decrease the ambiguity
in the value of $K_{\infty}$ as compared with the known estimate
$K_{\infty} = 210 \pm 30$ MeV.
However, our calculations confirm that an account of the QPC
is necessary to describe the total width of the ISGMR.

The work was supported by the Deutsche Forschungsgemeinschaft
under Grant No. 436~RUS~113/994/0-1,
by the Russian Foundation for Basic Research
under Grant No. 09-02-91352-DFG\_a,
and by the Hessian LOEWE initiative through the Helmholtz International
Center for FAIR.
V.~T. and E.~L. acknowledge financial support from the
Russian Federal Agency of Education under project No. 2.1.1/4779.

\end{document}